\documentclass[a4paper,11pt]{article}
\pdfoutput=1 

\usepackage{jheppub} 

\usepackage[T1]{fontenc}
\usepackage{multirow,bbold,slashed}
 
\allowdisplaybreaks
 
\definecolor{nicered}{rgb}{0.7,0.1,0.1} 
\definecolor{nicegreen}{rgb}{0.1,0.5,0.1}
\definecolor{niceblue}{rgb}{0.0,0.1,0.7}
\hypersetup{colorlinks,citecolor=niceblue,linkcolor=niceblue,urlcolor=niceblue}

\usepackage[normalem]{ulem}

\def \bm#1{\mbox{\boldmath$#1$\unboldmath}}
\def \beq{\begin{equation}}
\def \eeq{\end{equation}}
\def \bea{\begin{eqnarray}}
\def \eea{\end{eqnarray}}

\title{Precision tests of third-generation four-quark operators: $\bm{gg \to h}$ and $\bm{h \to \gamma \gamma}$}

\author[a]{Ulrich Haisch}
\author[a]{and Marco Niggetiedt}

\affiliation[a]{Max Planck Institute for Physics, \\ Boltzmannstr.~8, 85748 Garching, Germany}

\emailAdd{haisch@mpp.mpg.de, marco.niggetiedt@mpp.mpg.de}

\preprint{MPP-2025-147}

\abstract{We compute the two-loop contributions to Higgs production via gluon-gluon fusion ($gg \to h$) and Higgs decay into two photons ($h \to \gamma\gamma$), arising from third-generation four-quark operators in the Standard Model effective field theory (SMEFT). Our analysis is performed in the broken phase of the theory, retaining the full dependence on the Higgs and heavy-quark masses. This includes both finite matching corrections and logarithmic effects stemming from the renormalization group evolution within the SMEFT. As a byproduct, two-loop anomalous dimensions in the SMEFT are obtained. We also briefly discuss the phenomenological implications of our two-loop calculations.}

\begin{document} 
\maketitle
\flushbottom

\section{Introduction} 
\label{sec:introduction}

Run~2 of the Large Hadron Collider~(LHC) significantly improved our understanding of the properties of the Higgs boson. For example, both ATLAS and CMS~\cite{ATLAS:2022vkf,CMS:2022dwd} reduced the uncertainty in the measurement of the inclusive Higgs boson production cross section, which is dominated by the gluon-gluon fusion process~($gg \to h$), to approximately~$5\%$. Similarly, the signal strength for the Higgs decay to two photons~($h \to \gamma \gamma$) was measured with improved precision, reaching an accuracy of again about $5\%$. The improved understanding of the Higgs boson and its interactions enhances the sensitivity for indirectly probing physics beyond the Standard Model~(BSM). 

If new physics is characterized by heavy masses on the order of the scale $\Lambda$, the Standard~Model effective field theory~(SMEFT)~\cite{Buchmuller:1985jz,Grzadkowski:2010es,Brivio:2017vri,Isidori:2023pyp} provides a systematic and general framework for describing BSM phenomena at colliders. A set of dimension-six operators in the SMEFT, which is still weakly constrained, consists of four-quark operators that only involve third-generation fields. The weak constraints on the Wilson coefficients of these operators arise because such interactions can only be directly probed at tree level through processes like $t \bar t t \bar t$, $t \bar t b \bar b$, or $b \bar b b \bar b$ production. Although both ATLAS and CMS have observed evidence for four-top production~\cite{CMS:2019rvj,ATLAS:2020hpj,ATLAS:2021kqb,CMS:2023zdh,ATLAS:2023ajo}, the constraints on the~$t \bar t t \bar t$ cross section still exhibit large uncertainties of around $30\%$. Measurements of $t \bar t b \bar b$ production are also available~\cite{ATLAS:2018fwl,CMS:2019eih}, but their precision is currently limited as well. Furthermore, meaningful constraints on the Wilson coefficients for third-generation four-quark operators only arise when quadratic terms in the~$1/\Lambda^2$ expansion are included in global SMEFT analyses~\cite{Ethier:2021bye,Hartland:2019bjb,Degrande:2024mbg,Celada:2024mcf}, which raises concerns about the reliability of the SMEFT power counting. 

Due to these limitations, indirect tests for probing third-generation four-quark operators have attracted increasing attention~\cite{deBlas:2015aea,Gauld:2015lmb,Hartmann:2016pil,Boughezal:2019xpp,Brivio:2019ius,Degrande:2020evl,Dawson:2022bxd,Alasfar:2022zyr,DiNoi:2023ygk,Heinrich:2023rsd,Degrande:2024mbg,Haisch:2024wnw,DiNoi:2025uhu}. In this article, we calculate the two-loop contributions to the processes $gg \to h$ and $h \to \gamma \gamma$ arising from insertions of third-generation four-quark operators. We determine the exact dependence on the Higgs and heavy-quark masses, accounting for both finite matching corrections and the logarithmic effects due to the renormalization group~(RG) running in the SMEFT. It turns out that the renormalization of the relevant two-loop amplitudes requires certain two-loop SMEFT anomalous dimensions, which we calculate as a byproduct. Our computation allows us to revisit and cross-check several findings reported in~\cite{Alasfar:2022zyr,DiNoi:2023ygk,Heinrich:2023rsd,DiNoi:2025uhu}, which likewise investigate the impact of third-generation four-quark operators on Higgs physics observables.

This article is structured as follows: in~Section~\ref{sec:framework}, we specify the subset of SMEFT operators that are relevant to this study. A concise overview of the basics steps in calculating the two-loop SMEFT corrections to the~$gg \to h$ and $h \to \gamma \gamma$ processes is provided in~Section~\ref{sec:calculation}. Section~\ref{sec:ADMs} presents our findings for the two-loop anomalous dimensions relevant to this work, while Section~\ref{sec:matching} details the two-loop matching conditions for $gg \to h$ and $h \to \gamma \gamma$. In~Section~\ref{sec:pheno}, we study the phenomenological implications of our calculations. Our conclusions are presented in~Section~\ref{sec:conclusions}. Additional technical details on the choice of regularization and renormalization schemes, as~well~as the treatment of~$\gamma_5$, are provided in Appendices~\ref{app:mixren},~\ref{app:MSshifts}, and~\ref{app:HVresults}. Without further ado, let's crack straight into it. 

\section{Theoretical framework} 
\label{sec:framework}

To set our notation and conventions, we start by defining the SMEFT Lagrangian:
\beq \label{eq:LSMEFT}
{\cal L}_{\rm SMEFT} = \sum_i C_i (\mu) \hspace{0.25mm} Q_i \,.
\eeq
Here, $C_i (\mu)$ represents dimensionful Wilson coefficients evaluated at the renormalization scale $\mu$, which multiply the corresponding effective operators $Q_i$. Throughout this article, we assume that all Wilson coefficients are real.

In the Warsaw operator basis~\cite{Grzadkowski:2010es}, the complete set of dimension-six four-quark operators involving only third-generation fields consists of the following operators:
\beq \label{eq:4HFoperators} 
\begin{gathered}
Q^{(1)}_{qq} = (\bar q \gamma_\mu q) (\bar q \gamma^\mu q) \,, \qquad Q^{(3)}_{qq} = (\bar q \gamma_\mu \sigma^i q) (\bar q \gamma^\mu \sigma^i q) \,, \\[2mm]
Q^{(1)}_{qt} = (\bar q \gamma_\mu q) (\bar t \gamma^\mu t) \,, \qquad Q^{(8)}_{qt} = (\bar q \gamma_\mu T^a q)(\bar t \gamma^\mu T^a t) \,, \\[2mm]
Q^{(1)}_{qb} = (\bar q \gamma_\mu q) (\bar b \gamma^\mu b) \,, \qquad Q^{(8)}_{qb} = (\bar q \gamma_\mu T^a q)(\bar b \gamma^\mu T^a b) \,, \\[2mm]
Q^{(1)}_{tt} = (\bar t \gamma_\mu t)(\bar t \gamma^\mu t) \,, \qquad  Q^{(1)}_{bb} = (\bar b \gamma_\mu b)(\bar b \gamma^\mu b) \,, \\[2mm]
Q^{(1)}_{tb} = (\bar t \gamma_\mu t)(\bar b \gamma^\mu b) \,, \qquad Q^{(8)}_{tb} = (\bar t \gamma_\mu T^a t)(\bar b \gamma^\mu T^a b) \,, \\[2mm]
Q^{(1)}_{qtqb} = (\bar q t) \hspace{0.5mm} \varepsilon \hspace{0.5mm} (\bar q b) \,, \qquad Q^{(8)}_{qtqb} = (\bar q T^a t) \hspace{0.5mm} \varepsilon \hspace{0.5mm} (\bar q T^a b) \,. 
\end{gathered}
\eeq
The symbol $q$ represents the left-handed third-generation quark $SU(2)_L$ doublets, while $t$ and $b$ denote the right-handed top-quark and bottom-quark $SU(2)_L$ singlets, respectively. The~$\sigma^i$ are the Pauli matrices, and $T^a =\lambda^a/2$ are the~$SU(3)_C$ generators, where $\lambda^a$ are the Gell-Mann matrices. Additionally, $\varepsilon = i \sigma^2$ is the antisymmetric~$SU(2)$ tensor. Note that for~$Q^{(1)}_{qtqb}$ and~$Q^{(8)}_{qtqb}$, the sum of the Hermitian conjugate is implied in~(\ref{eq:LSMEFT}).

In this article, we calculate the two-loop contributions of the operators~(\ref{eq:4HFoperators}) to the processes $gg \to h$ and $h \to \gamma \gamma$. Within the Warsaw operator basis of the SMEFT, these transitions receive tree-level contributions from the following dimension-six operators
\beq \label{eq:Higgsoperators}
\begin{gathered}
Q_{HG} = H^\dagger H \hspace{0.25mm} G_{\mu \nu}^a G^{a, \mu \nu} \,, \qquad Q_{HB} = H^\dagger H \hspace{0.25mm} B_{\mu \nu} B^{\mu \nu} \,, \\[2mm]
Q_{HW} = H^\dagger H \hspace{0.25mm} W_{\mu \nu}^i W^{i, \mu \nu} \,, \qquad Q_{HW\!B} = H^\dagger \sigma^i H \hspace{0.25mm} W^{i}_{\mu\nu} B^{\mu\nu} \,, 
\end{gathered}
\eeq
where $H$ represents the Standard Model~(SM) Higgs doublet, while $G_{\mu \nu}^a$, $B_{\mu \nu}$, and $W_{\mu \nu}^i$ correspond to the field strength tensors of the~$SU(3)_C$, $U(1)_Y$, and $SU(2)_L$ gauge fields, respectively. The corresponding gauge couplings will be denoted by $g_s$, $g_1$, and $g_2$, respectively. 

To renormalize the two-loop $gg \to h$ and $h \to \gamma \gamma$ amplitudes with insertions of third-generation four-quark operators, it is also necessary to account for the following operators 
\beq \label{eq:renoperators}
\begin{gathered}
Q_{tH} = H^\dagger H \hspace{0.25mm} \bar q \hspace{0.25mm} \widetilde H \hspace{0.25mm} t \,, \qquad Q_{bH} = H^\dagger H \hspace{0.25mm} \bar q \hspace{0.25mm} H \hspace{0.25mm} b \,, \\[2mm]
Q_{tG} = \bar q \hspace{0.25mm} \sigma^{\mu \nu} \widetilde H \hspace{0.25mm} T^a t \hspace{0.5mm} G_{\mu \nu}^a \,, \qquad Q_{tB} = \bar q \hspace{0.25mm} \sigma^{\mu \nu} \widetilde H \hspace{0.25mm} t \hspace{0.5mm} B_{\mu \nu} \,, \qquad Q_{tW} = \bar q \hspace{0.25mm} \sigma^{\mu \nu} \widetilde H \hspace{0.25mm} \sigma^i \hspace{0.25mm} t \hspace{0.5mm} W_{\mu \nu}^i \,,\\[2mm]
Q_{bG} = \bar q \hspace{0.25mm} \sigma^{\mu \nu} H \hspace{0.25mm} T^a b \hspace{0.5mm} G_{\mu \nu}^a \,, \qquad Q_{bB} = \bar q \hspace{0.25mm} \sigma^{\mu \nu} H \hspace{0.25mm} b \hspace{0.5mm} B_{\mu \nu} \,, \qquad Q_{bW} = \bar q \hspace{0.25mm} \sigma^{\mu \nu} H \hspace{0.25mm} \sigma^i \hspace{0.25mm} b \hspace{0.5mm} W_{\mu \nu}^i \,,\
\end{gathered}
\eeq
where we use the shorthand notation $\widetilde H = \varepsilon \cdot H^\ast$, with $\varepsilon_{12} = +1$, and where, for all these operators, the Hermitian conjugate term in~(\ref{eq:LSMEFT}) is implicitly included. These operators are required to cancel ultraviolet~(UV) singularities that arise as one-loop subdivergences within the two-loop amplitudes. Our conventions for covariant derivatives acting on fermion fields, field strength tensors, and other field-theoretic entities follow those adopted in~\cite{Grzadkowski:2010es}. For further details, we refer the interested reader to this work.

\section{Calculation in a nutshell} 
\label{sec:calculation}

In our two-loop calculations we employ dimensional regularization to handle UV and infrared (IR) singularities in~$d = 4 - 2 \hspace{0.125mm} \epsilon$ supplemented by the renormalisation scale~$\mu$ and a~naive anti-commuting~$\gamma_5$~(NDR)~\cite{Chanowitz:1979zu}. We have verified that the NDR scheme yields consistent results for all observables considered, as non-zero traces involving $\gamma_5$ do not contribute after integrating over loop momenta. The~actual computation made use of the {\tt Mathematica} packages {\tt FeynRules}~\cite{Alloul:2013bka}, {\tt FeynArts}~\cite{Hahn:2000kx}, {\tt FormCalc}~\cite{Hahn:2016ebn}, and {\tt LiteRed}~\cite{Lee:2013mka}. Specifically, {\tt FeynRules} is used to implement the SMEFT operators~(\ref{eq:4HFoperators}) to~(\ref{eq:renoperators}) and generate {\tt FeynArts} model~files. {\tt FeynArts} is then employed to construct the relevant Feynman diagrams and amplitudes, while {\tt FormCalc} performs the projection onto form factors (see, for instance,~\cite{Steinhauser:2002rq}) and handles the color and Dirac algebra. The scalar integrals are then reduced to master integrals using {\tt LiteRed}. Notably, the master integrals consist exclusively of massive one-loop bubble and triangle integrals, whose $\epsilon$-expansion is well known in terms~of harmonic polylogarithms~(see, for example, \cite{Anastasiou:2006hc}). The same chain of {\tt Mathematica} packages was employed in the recent works~\cite{Haisch:2025lvd,Haisch:2025pql}. The reduction to master integrals and their evaluation have been independently verified using an in-house software package, yielding identical final results.

\begin{figure}[t!]
\begin{center}
\includegraphics[width=0.95\textwidth]{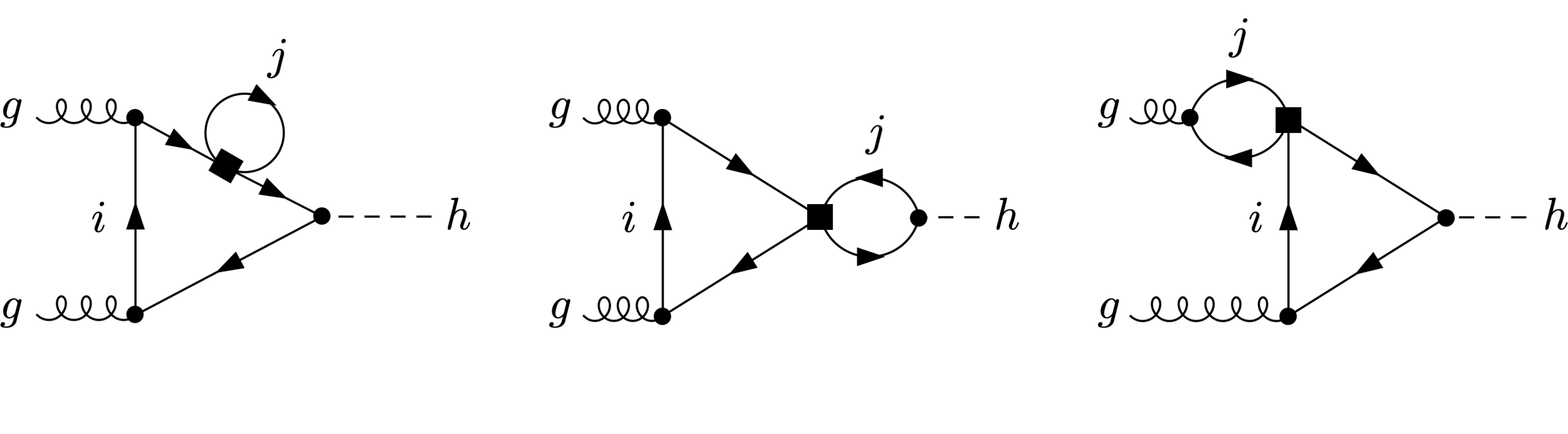}
\end{center}
\vspace{-6mm} 
\caption{\label{fig:diagrams} Examples of two-loop contributions to the~$gg \to h$ process involving insertions of one of the dimension-six operators~(black~squares) introduced in~(\ref{eq:4HFoperators}). The internal quarks $i$ and $j$ can be both a top quark~($t$) or a bottom quark~($b$). Consult the main text for further explanations.} 
\end{figure}

Figure~\ref{fig:diagrams} displays representative two-loop Feynman diagrams with insertions of third-generation four-quark operators contributing to the~$gg \to h$ process. The bare amplitude contains both $1/\epsilon^2$ and $1/\epsilon$ poles of UV origin, with no associated IR divergences. The~absence of IR singularities is straightforward to understand, as no real emission processes involving the dimension-six operators in~(\ref{eq:4HFoperators}) contribute at the relevant perturbative order to inclusive $gg \to h$ production. As shown in the figure, the amplitude receives contributions from three distinct topologies. The first class of topologies, exemplified by the diagram on the left-hand side of the figure, involves the insertion of one of the operators in~(\ref{eq:4HFoperators}) into an internal propagator. The contributions from these diagrams are precisely canceled by the corresponding mass counterterms when the heavy-quark masses are renormalized in the on-shell~(OS) scheme. The relevant mass counterterms can easily be derived from the results presented in~\cite{Gauld:2015lmb}, and we agree with these findings. 

A representative example of the second type of topology is displayed in the center of Figure~\ref{fig:diagrams}. These contributions become UV finite after renormalizing the Yukawa couplings~$y_t$ and $y_b$ by considering the mixing of certain third-generation four-quark operators~(\ref{eq:4HFoperators}) into the Yukawa-type operators $Q_{tH}$ and $Q_{bH}$ introduced in~(\ref{eq:4HFoperators}). In the main part of our work, we renormalize the heavy-quark Yukawa couplings using the same scheme as the heavy-quark masses, which means we adopt the~OS~scheme.\footnote{The analytical expressions for the finite shifts of the two-loop $gg \to h$ and $h \to \gamma \gamma$ amplitudes that arise from changing the renormalization scheme from OS~to~$\overline{\rm MS}$ are provided in~Appendix~\ref{app:MSshifts}.} Operator renormalization is instead carried out in the modified minimal subtraction ($\overline{\rm MS}$) scheme. We have verified that after renormalizing the~$y_t$ and $y_b$ couplings, the~$1/\epsilon$ structure of the second type of contributions is proportional to the one-loop anomalous dimensions for $Q_{tH}$ and $Q_{bH}$ computed first in the publications~\cite{Jenkins:2013zja,Jenkins:2013wua}. Further details on the implementation of the mixed~$\overline{\rm MS\hspace{0.25mm}}$–${\rm OS}$ renormalization scheme in SMEFT calculations can be found in Appendix~\ref{app:mixren} as~well~as in the publications~\cite{Gauld:2015lmb,Gauld:2016kuu,Dawson:2018pyl,Cullen:2019nnr,Cullen:2020zof,Alasfar:2022zyr,Heinrich:2023rsd,Haisch:2024wnw,DiNoi:2025uhu}.

The Feynman diagram on the right-hand side of~Figure~\ref{fig:diagrams} illustrates an example of the third type of topology contributing to the~$gg \to h$ process. We find that the bare amplitudes with identical internal quarks contain only local $1/\epsilon$ poles of UV origin, while the matrix elements involving a combination of top and bottom quarks exhibit $1/\epsilon^2$ divergences as well. In the first case, the UV singularities are canceled by adding two-loop counterterms that account for the mixing of the operators $Q_{qi}^{(1)}$ and $Q_{qi}^{(8)}$ into the operator $Q_{HG}$ introduced in~(\ref{eq:Higgsoperators}). For the case of $Q_{qt}^{(1)}$ and $Q_{qt}^{(8)}$ this observation has been made already in~\cite{DiNoi:2023ygk} and~\cite{Heinrich:2023rsd} in the context of $gg \to h$ and $gg \to hh$ production, respectively. In the case of the mixed top and bottom quark contributions, the locality of UV divergences requires that the structure of the~$1/\epsilon^2$ pole, which is proportional to the product of $y_t \hspace{0.125mm} y_b$, is related to products of one-loop~$1/\epsilon$ subdivergences~(for general discussions of this point see, for~instance, \cite{Chetyrkin:1997fm,Gambino:2003zm}). The relevant operators are $Q_{qtqb}^{(1)}$ and $Q_{qtqb}^{(8)}$, which mix at one-loop into the chromomagnetic dipole operators $Q_{tG}$ and $Q_{bG}$ defined in~(\ref{eq:renoperators}), and these operators, in turn, mix at one-loop into $Q_{HG}$. All~relevant one-loop anomalous dimensions can be found in~\cite{Alonso:2013hga}, and after subtraction of the relevant counterterm contributions, no $1/\epsilon$ pole remains. Consequently, no additional two-loop counterterm proportional to $y_t \hspace{0.125mm} y_b$ is needed to account for the mixing of $Q_{qtqb}^{(1)}$ and $Q_{qtqb}^{(8)}$ into $Q_{HG}$ to make the corresponding amplitudes UV finite. 

The calculation and renormalization of the two-loop amplitudes for $h \to \gamma \gamma$ proceed in complete analogy to the~$gg \to h$ case discussed above. The only difference is that, in this case, one must consider the three operators $Q_{HB}$, $Q_{HW}$, and $Q_{HW\!B}$ instead of the single operator $Q_{HG}$, as~well~as the four electroweak~(EW) dipole operators $Q_{tB}$, $Q_{tW}$, $Q_{bB}$, and $Q_{bW}$ instead of the two chromomagnetic dipole operators $Q_{tG}$ and $Q_{bG}$. The~one-loop anomalous dimensions relevant for operator renormalization are all provided in the articles~\cite{Jenkins:2013zja,Jenkins:2013wua,Alonso:2013hga}, and we have verified that their inclusion renders the remaining $1/\epsilon$~poles in the two-loop $h \to \gamma \gamma$ amplitudes either local or removes them. The left-over UV divergences determine the two-loop anomalous dimensions that describe the mixing of~(\ref{eq:4HFoperators}) into~$Q_{HB}$, $Q_{HW}$, and~$Q_{HW\!B}$.

\section{Two-loop anomalous dimensions} 
\label{sec:ADMs}

As explained in the previous section, the local $1/\epsilon$ poles that remain in our two-loop computations of the~$gg \to h$ and $h \to \gamma \gamma$ amplitudes after one-loop renormalization are of UV origin, and can therefore be used to extract  two-loop anomalous dimensions in the SMEFT.\footnote{Other contributions to the two-loop beta functions in the SMEFT have been reported in~\cite{Gorbahn:2016uoy,Bern:2020ikv,Jin:2020pwh,Haisch:2022nwz,DiNoi:2023ygk,Jenkins:2023bls,DiNoi:2024ajj,Born:2024mgz,Naterop:2024cfx,Duhr:2025zqw,Haisch:2025lvd,Zhang:2025ywe,Assi:2025fsm,Naterop:2025cwg,DiNoi:2025arz,Haisch:2025vqj,Duhr:2025yor,DiNoi:2025tka,Banik:2025wpi,Henriksson:2025vyi,Chala:2025crd,Guedes:2025sax,Born:2026xkr,Duhr:2026btw}.} The RG evolution of the Wilson coefficients $C_i$ is dictated by the beta functions:
\beq \label{eq:betai}
\frac{d C_i}{d \ln \mu} = \beta_i = \sum_{l=1}^\infty \frac{\beta_i^{(l)}}{(16 \pi^2)^l} \,. 
\eeq
The one-loop beta functions $\beta_i^{(1)}$ for the operators in~(\ref{eq:Higgsoperators}), describing the mixing of the operators in~(\ref{eq:4HFoperators}) into those in~(\ref{eq:Higgsoperators}), vanish identically. In contrast, within the NDR~scheme, we obtain the following expressions for the two-loop beta functions
\beq \label{eq:beta2} 
\begin{gathered}
\beta_{HG}^{(2)} = -8 \hspace{0.25mm} T_F \hspace{0.25mm} g_s^2 \hspace{0.25mm} \sum_{i=t,b} y_i^2 \left [ C_{qi}^{(1)} + \left ( C_F - \frac{C_A}{2} \right ) C_{qi}^{(8)} \right ] \,, \\[2mm] 
\beta_{HB}^{(2)} = -2 \hspace{0.25mm} C_A \hspace{0.25mm} g_1^2 \sum_{i=t,b} \left ( Y_{L, i} + Y_{R, i} \right )^2 \hspace{0.25mm} y_i^2 \left ( C_{qi}^{(1)} + C_F \hspace{0.25mm} C_{qi}^{(8)} \right ) \,, \\[2mm]
\beta_{HW}^{(2)} = -2 \hspace{0.25mm} C_A \hspace{0.25mm} g_2^2 \sum_{i=t,b} T_{3,i}^2 \hspace{0.5mm} y_i^2 \left ( C_{qi}^{(1)} + C_F \hspace{0.25mm} C_{qi}^{(8)} \right ) \,, \\[2mm]
\beta_{HW\!B}^{(2)} = 4 \hspace{0.25mm} C_A \hspace{0.25mm} g_1\hspace{0.125mm} g_2 \sum_{i=t,b} \left ( Y_{L, i} + Y_{R, i} \right ) \hspace{0.25mm} T_{3,i} \hspace{0.5mm} y_i^2 \left ( C_{qi}^{(1)} + C_F \hspace{0.25mm} C_{qi}^{(8)} \right ) \,,
\end{gathered}
\eeq
which can be obtained from the computations presented in this work. Above the dependence of the gauge and Yukawa couplings as~well~as the Wilson coefficients on the renormalization scale $\mu$ has been omitted. In~(\ref{eq:beta2}), $y_i = \sqrt{2} m_i/v$, with $m_i$ being the heavy-quark masses and $v = 246.22 \, {\rm GeV}$ is the Higgs vacuum expectation value. Additionally, $T_F = 1/2$ represents the trace normalization factor of the~$SU(3)_C$ generators, i.e.~${\rm tr} \left ( T^a T^b \right ) = T_F \hspace{0.25mm} \delta^{ab}$, while the quadratic Casimir operators of $SU(3)_C$ are $C_A = 3$ and $C_F = 4/3$. Our results for the $C_{qt}^{(1)}$ and $C_{qt}^{(8)}$ contributions to $\beta_{HG}^{(2)}$ agree with the two-loop NDR results reported in~\cite{DiNoi:2023ygk,Heinrich:2023rsd,DiNoi:2024ajj}. They are also consistent with the two-loop beta functions for the complete set of dimension-six SMEFT operators recently presented in~\cite{Born:2026xkr}, after transforming from the Warsaw to the Mainz operator basis. Likewise, our results for $\beta_{HB}^{(2)}$, $\beta_{HW}^{(2)}$, and $\beta_{HWB}^{(2)}$ agree with those of~\cite{Born:2026xkr} upon performing the corresponding basis transformation. In these expressions, $Y_{L,i}$ and $Y_{R,i}$ denote the hypercharges of the left- and right-handed internal heavy quarks, respectively, with $Y_{L,t}=Y_{L,b}=1/6$, $Y_{R,t}=2/3$, and $Y_{R,b}=-1/3$, while $T_{3,i}$~denotes the corresponding third component of weak isospin, with $T_{3,t}=-T_{3,b}=1/2$.

We note that the results in~(\ref{eq:beta2}) are scheme-dependent, a point previously stressed in~\cite{DiNoi:2023ygk,Heinrich:2023rsd,DiNoi:2024ajj,DiNoi:2025uhu} for the contributions of the Wilson coefficients $C_{qt}^{(1)}$ and $C_{qt}^{(8)}$ to the beta function $\beta_{HG}^{(2)}$.\footnote{The results for the two-loop beta functions~(\ref{eq:beta2}) in the ’t Hooft-Veltman (HV) scheme~\cite{tHooft:1972tcz,Breitenlohner:1977hr} are provided in~Appendix~\ref{app:HVresults}.} This dependence on the regularization scheme is a distinctive feature linked to the mixing of the operators in~(\ref{eq:4HFoperators}) with different chiralities into the dipole operators in~(\ref{eq:renoperators}), which are also chirality-flipping. It is reminiscent of the same subtlety discussed over 30 years ago in the context of the~$b \to s \gamma$ and $b \to s g$ decays~\cite{Ciuchini:1993ks,Buras:1993xp,Ciuchini:1993fk}. The scheme dependence of~(\ref{eq:beta2}) is an inherent and unavoidable aspect of next-to-leading order~(NLO) SMEFT beta functions. In the present case, the scheme dependence originates from the treatment of $\gamma_5$. However, it can for instance also arise from the choice of physical and evanescent operators in the SMEFT, as discussed in some detail in~\cite{Degrande:2024mbg,Haisch:2024wnw}. A properly chosen regularization and renormalization scheme, along with a suitable choice of operator basis, can thus aid in simplifying SMEFT computations beyond the leading order. While~we do not advocate for a specific approach here, we just point out that the NDR scheme, combined with a suitable choice of four-quark operators~\cite{Chetyrkin:1997gb}, enabled the calculation of the three-loop and four-loop anomalous dimensions for $b \to s l^+ l^-$~\cite{Gorbahn:2004my} and~$b \to s \gamma$~\cite{Czakon:2006ss}, respectively. Applying strategies akin to those used in~\cite{Gorbahn:2004my,Czakon:2006ss} to multiloop calculations within the SMEFT framework therefore does not seem like an entirely unreasonable idea.

\section{Two-loop matching conditions} 
\label{sec:matching}

We now present the two-loop matching corrections to the $gg \to h$ scattering amplitudes in the SMEFT, expressed in terms of the corresponding form factors and involving the Wilson coefficients of the third-generation four-quark operators~(\ref{eq:4HFoperators}). To establish our notation and normalization, we first consider the one-loop contribution to the $gg \to h$ scattering amplitude in the SM induced by heavy quarks. We write the amplitude in terms of the corresponding form factor as
\beq \label{eq:gghamplitude}
\mathcal{M}^{ab}_{\mu\nu} = \frac{\alpha_s}{3 \pi v} \hspace{0.5mm} \delta^{ab}  \left( g_{\mu\nu}\,p_1\!\cdot p_2-p_{1\hspace{0.125mm}\nu} \hspace{0.25mm} p_{2\hspace{0.125mm}\mu} \right) G^{(1)} \,, \qquad  G^{(1)} = \sum_{i=t,b} G^{(1)}_i \,, 
\eeq
where $\alpha_s = g_s^2/(4\pi)$ denotes the QCD coupling, while $p_1$ and $p_2$ are the momenta of the incoming gluons, carrying Lorentz indices $\mu$ and $\nu$ and colour indices $a$ and $b$, respectively. The one-loop form factor entering~(\ref{eq:gghamplitude}) is given by
\beq \label{eq:gghoneloop}
G_i^{(1)} = 3 \hspace{0.25mm} T_F \hspace{0.5mm} \Big [ \tau_i + 2 \left ( \tau_i - 1 \right ) F_{\rm tri} (x_i) \Big ] \,,
\eeq
with 
\beq \label{eq:xitauiFtri}
x_i = \frac{\sqrt{1 - \tau_i} - 1}{\sqrt{1 - \tau_i} + 1} + i \varepsilon \,, \qquad \tau_i = \frac{4 m_i^2}{m_h^2} \,, \qquad F_{\rm tri} (x_i) = \frac{\tau_i}{8} \ln^2 x_i \,, 
\eeq
and $\varepsilon> 0$ is infinitesimal. In~(\ref{eq:gghoneloop}), and $F_{\rm tri} (x_i)$ denotes the~$\epsilon^0$ contribution to the massive one-loop triangle integral, adopting the notation of~\cite{Anastasiou:2006hc}. Observe that for $\tau_i \to \infty$, we have $G_i^{(1)} \to 2 \hspace{0.25mm} T_F$, which establishes the normalization of the~$gg \to h$ form~factor.

The two-loop corrections to the~$gg \to h$ form factor arising from a single insertion of the operators in~(\ref{eq:4HFoperators}), which contribute additively to $G^{(1)}$ as defined in~(\ref{eq:gghamplitude}), can be decomposed into four distinct flavor combinations
\beq \label{eq:gghtwoloop}
G^{(2)} = \frac{m_h^2}{16 \pi^2} \sum_{ij = tt, tb, bt, bb } G_{ij}^{(2)} \,, 
\eeq
where the first~(second) subscript of $G_{ij}^{(2)}$ indicates the flavor of the quark within the triangle~(bubble) subdiagram in the Feynman graphs shown in Figure~\ref{fig:diagrams}. Our results for the functions $G_{ij}^{(2)}$ in the mixed~$\overline{\rm MS\hspace{0.25mm}}$-$\hspace{0.125mm}{\rm OS}$ renormalization scheme with an NDR treatment of~$\gamma_5$~are\footnote{The adjustments to the form factors~(\ref{eq:Gij2}) and~(\ref{eq:Aij2}) in the HV scheme are discussed in~Appendix~\ref{app:HVresults}.} 
\bea \label{eq:Gij2}
\begin{split}
& \hspace{1.5cm} G_{ii}^{(2)} = \left ( \tau_i - 1 \right ) \Big ( 1 + 2 F_{\rm bub} (x_i) + 2 L_i \Big ) \hspace{0.5mm} G_i^{(1)} \left ( C_{qi}^{(1)} + C_F \hspace{0.25mm} C_{qi}^{(8)} \right ) \\[2mm]
& \hspace{1.75cm} + 3 \hspace{0.25mm} T_F \hspace{0.25mm} \tau_i \hspace{0.5mm} \Big ( F_{\rm bub} (x_i) + 2 F_{\rm tri} (x_i) + 2 L_i \Big ) \left [ C_{qi}^{(1)} + \left ( C_F - \frac{C_A}{2} \right ) C_{qi}^{(8)} \right ] \,, \\[4mm]
& G_{ij}^{(2)} = \frac{m_j}{2 \hspace{0.25mm} m_i} \left ( \tau_j - 1 \right ) \Big ( 1 + F_{\rm bub} (x_j) + L_j \Big ) \hspace{0.5mm} G_i^{(1)} \left [ \hspace{0.25mm} \big ( 2 C_A + 1 \big) \hspace{0.25mm} C_{qtqb}^{(1)} + C_F \hspace{0.25mm} C_{qtqb}^{(8)} \right ] \\[2mm]
& \hspace{0.25cm} + \frac{m_i \hspace{0.125mm} m_j}{2 \hspace{0.25mm} m_h^2} \hspace{0.5mm} T_F \hspace{0.5mm} \Big \{ \hspace{0.25mm} 12 \hspace{0.5mm} \big [ F_{\rm bub} (x_i) + 2 F_{\rm tri} (x_i) \big ] \hspace{0.5mm} L_j + 12 L_i L_j \Big \} \left [ C_{qtqb}^{(1)} + \left ( C_F - \frac{C_A}{2} \right ) C_{qtqb}^{(8)} \right ] \,, 
\end{split}
\eea
where the first~(second) term in both the formulas of $G_{ii}^{(2)}$ and $G_{ij}^{(2)}$ corresponds to contributions of Feynman diagrams of the second~(third) type. The analytic expression for $F_{\rm tri} (x_i)$ is provided in~(\ref{eq:xitauiFtri}), while $F_{\rm bub} (x_i)$, in the notation of~\cite{Anastasiou:2006hc}, represents the~$\epsilon^0$ term of the massive one-loop bubble integral. Specifically, 
\beq \label{eq:Fbub}
F_{\rm bub} (x_i) = 1 + \frac{1 + x_i}{1 - x_i} \hspace{0.5mm} \ln x_i \,,
\eeq 
and we have introduced the abbreviation $L_i = \ln \left ( \mu^2/m_i^2 \right )$. Note that since~(\ref{eq:Gij2}) corresponds to the mixed~$\overline{\rm MS\hspace{0.25mm}}$-$\hspace{0.125mm}{\rm OS}$ scheme for the renormalization of heavy-quark masses and Yukawa couplings, the logarithms $L_i$ in the above functions $G_{ij}^{(2)}$ are linked to the RG flow of the operators~(\ref{eq:4HFoperators}) into~$Q_{HG}$. Before shifting our focus to the~$h \to \gamma \gamma$ decay, we mention that the above result for $G_{tt}^{(2)}$ is consistent with the corresponding parts of the two-loop $gg \to h$ and $gg \to hh$ amplitudes computed in~\cite{DiNoi:2023ygk,DiNoi:2025uhu} and~\cite{Heinrich:2023rsd}, respectively, when translated to the same scheme. However, the results in~(\ref{eq:Gij2}) differ from the expressions reported in~\cite{Alasfar:2022zyr}. For~instance, the contributions that result from the third class of graphs shown in~Figure~\ref{fig:diagrams} seem to be entirely missing in formula~(14) of that work. Given the lack of detailed information in~\cite{Alasfar:2022zyr} regarding the regularization scheme, including the treatment of~$\gamma_5$, we are unable to provide further clarification on the observed discrepancies.

In the case of the~$h \to \gamma \gamma$ decay, the full one-loop contribution to the scattering amplitude in the SM can be written as
\beq \label{eq:haaamplitude}
\mathcal{M}_{\mu\nu} = \frac{\alpha}{2 \pi v}  \left ( g_{\mu\nu} \, p_1 \cdot p_2-p_{1\hspace{0.125mm}\nu}\hspace{0.25mm}p_{2\hspace{0.125mm}\mu} \right ) A^{(1)} \,, \qquad  A^{(1)} = \sum_{i=t,b,\tau} A_i^{(1)} +  A_W^{(1)}
\eeq
where $\alpha=e^2/(4\pi)$ denotes the electromagnetic fine-structure constant, and $p_1$ and $p_2$ are the momenta of the incoming photons with Lorentz indices $\mu$ and $\nu$, respectively. The~one-loop form factors~$A_i^{(1)}$ and $A_W^{(1)}$ entering~(\ref{eq:haaamplitude}) are given by
\beq \label{eq:1loopffdecay}
\begin{split}
A_i^{(1)} & = 2 \hspace{0.25mm} C_A \hspace{0.125mm} Q_i^2 \hspace{0.5mm} \Big [ \tau_i + 2 \left ( \tau_i - 1 \right ) F_{\rm tri} (x_i) \Big ] \,, \\[2mm] 
A_W^{(1)} & = -2 - 3 \hspace{0.25mm} \tau_W - 6 \left ( \tau_W - 2 \right ) F_{\rm tri} (x_W) \,.
\end{split}
\eeq
Here, $Q_i$ denotes the electromagnetic charge of the internal heavy quark in units of the elementary charge~$e$, with $Q_t = 2/3$ and $Q_b=-1/3$, while $\tau_W$ and $x_W$ are defined analogously to~(\ref{eq:xitauiFtri}). The form factor $A_W^{(1)}$ receives contributions from both virtual $W$ bosons and would-be Goldstone bosons in non-unitary gauges. For an internal tau lepton, the form factor $A_i^{(1)}$ can be used after replacing $C_A$ by $1$ and setting $Q_\tau=-1$. In the heavy-mass limits $\tau_i \to \infty$ and $\tau_W \to \infty$, the corresponding one-loop form factors approach $A_i^{(1)} \to 4/3 \hspace{0.25mm} C_A \hspace{0.125mm} Q_i^2$ and $A_W^{(1)} \to -7$, respectively. These limits fix the normalization of the~$h \to \gamma \gamma$ form factors.

The two-loop contributions to the~$h \to \gamma \gamma$ form factor from a single insertion of the operators in~(\ref{eq:4HFoperators}), which contribute additively to $A^{(1)}$ defined in~(\ref{eq:haaamplitude}), can be written as follows
\beq \label{eq:hgammagammatwoloop}
A^{(2)} = \frac{m_h^2}{16 \pi^2} \sum_{ij = tt, tb, bt, bb } A_{ij}^{(2)} \,,
\eeq
where, in the mixed~$\overline{\rm MS\hspace{0.25mm}}$-$\hspace{0.125mm}{\rm OS}$ renormalization scheme with an NDR treatment of~$\gamma_5$, the coefficients $A_{ij}^{(2)}$ are given by
\bea \label{eq:Aij2}
\begin{split}
& \hspace{1.20cm} A_{ii}^{(2)} = \left ( \tau_i - 1 \right ) \Big ( 1 + 2 F_{\rm bub} (x_i) + 2 L_i \Big ) \hspace{0.5mm} A_i^{(1)} \left ( C_{qi}^{(1)} + C_F \hspace{0.25mm} C_{qi}^{(8)} \right ) \\[2mm]
& \hspace{2.30cm} + 2 \hspace{0.25mm} C_A \hspace{0.125mm} Q_i^2 \hspace{0.5mm} \tau_i \hspace{0.5mm} \Big ( F_{\rm bub} (x_i) + 2 F_{\rm tri} (x_i) + 2 L_i \Big ) \left ( C_{qi}^{(1)} + C_F \hspace{0.25mm} C_{qi}^{(8)} \right ) \,, \\[4mm]
& A_{ij}^{(2)} = \frac{m_j}{2 \hspace{0.25mm} m_i} \left ( \tau_j - 1 \right ) \Big ( 1 + F_{\rm bub} (x_j) + L_j \Big ) \hspace{0.5mm} A_i^{(1)} \left [ \hspace{0.25mm} \big ( 2 C_A + 1 \big) \hspace{0.25mm} C_{qtqb}^{(1)} + C_F \hspace{0.25mm} C_{qtqb}^{(8)} \right ] \\[2mm]
& \hspace{1.1cm} + \frac{m_i \hspace{0.125mm} m_j}{m_h^2} \hspace{0.5mm} Q_i \hspace{0.25mm} Q_j \hspace{0.5mm} \Big \{ \hspace{0.25mm} 12 \hspace{0.5mm} \big [ F_{\rm bub} (x_i) + 2 F_{\rm tri} (x_i) \big ] \hspace{0.5mm} L_j + 12 L_i L_j \Big \} \left ( C_{qtqb}^{(1)} + C_F \hspace{0.25mm} C_{qtqb}^{(8)} \right ) \,.
\end{split}
\eea
Here, $A_i^{(1)}$ refers to the form factor~(\ref{eq:1loopffdecay}), which describes the one-loop corrections arising from heavy quarks in the SM. Note that in both expressions for $A_{ii}^{(2)}$ and $A_{ij}^{(2)}$, the first (second) term corresponds to the contributions from Feynman diagrams of the second~(third) type. The logarithms $L_i$ in~(\ref{eq:Aij2}) again stem from the RG running of the operators~(\ref{eq:4HFoperators}) into the pure EW operators entering~(\ref{eq:Higgsoperators}), because the above expressions for $A_{ij}^{(2)}$ employ the mixed~$\overline{\rm MS\hspace{0.25mm}}$-$\hspace{0.125mm}{\rm OS}$ scheme for the renormalization of heavy-quark masses and Yukawa couplings. The expression for $A_{tt}^{(2)}$ given above agrees with the corresponding terms in the two-loop calculation of $h \to \gamma \gamma$ presented very recently in~\cite{DiNoi:2025uhu}. However, the results in~(\ref{eq:Aij2}) do not match the expressions found in~\cite{Alasfar:2022zyr}.

\section{Phenomenological implications}
\label{sec:pheno}

Using the definitions and results given in~(\ref{eq:gghamplitude}) to~(\ref{eq:gghtwoloop}), the Higgs production cross section in gluon-gluon fusion, incorporating the effects of third-generation four-quark operators, can be expressed~as
\beq \label{eq:higgsXs}
\sigma \left (g g \to h \right ) = \frac{\alpha_s^2}{576 \hspace{0.125mm} \pi} \hspace{0.25mm} \frac{1}{v^2} \hspace{0.5mm} \big | G^{(1)} + G^{(2)} \big |^2 \,.
\eeq
To first order in the Wilson coefficients, the corresponding signal strength modification relative to the SM is therefore given by
\beq \label{eq:kappag}
\delta \mu_{gg} = 2 \hspace{0.125mm} \delta \kappa_g = \frac{2 \hspace{0.25mm} {\rm Re} \left ( G^{(1)} G^{(2) \hspace{0.25mm} \ast} \right )}{ \big | G^{(1)} \big |^2} \,. 
\eeq
For the Higgs decay into two photons, the corresponding formulas read 
\beq \label{eq:higgsdecay} 
\Gamma \left ( h \to \gamma \gamma \right) = \frac{\alpha^2}{256 \hspace{0.125mm} \pi^3} \hspace{0.25mm} \frac{m_h^3}{v^2} \hspace{0.5mm} \big | A^{(1)} + A^{(2)} \big |^2 \,, 
\eeq
and 
\beq \label{eq:dkappagamma}
\delta \mu_{\gamma \gamma} = 2 \hspace{0.125mm} \delta \kappa_\gamma = \frac{2 \hspace{0.25mm} {\rm Re} \left ( A^{(1)} A^{(2) \hspace{0.25mm} \ast} \right )}{ \big | A^{(1)} \big |^2} \,,
\eeq
with the relevant quantities entering $A^{(1)}$ and $A^{(2)}$ given in~(\ref{eq:haaamplitude}) to (\ref{eq:Aij2}).

By using the~OS~masses $m_t = 172.4 \, {\rm GeV}$, $m_b = 4.78 \, {\rm GeV}$, $m_\tau = 1.777 \, {\rm GeV}$, and $m_h = 125.2 \, {\rm GeV}$ from the Particle Data Group~\cite{ParticleDataGroup:2024cfk} as input parameters, and setting the renormalization scale to $\mu = m_h$, we obtain the following numerical results: 
\bea \label{eq:kappaggammanumerics}
\begin{gathered}
\frac{\delta \kappa_g}{v^2} = -0.085 \hspace{0.25mm} C_{qt}^{(1)} - 0.022 \hspace{0.25mm} C_{qt}^{(8)} + 0.002 \hspace{0.25mm} C_{qb}^{(1)} + 0.002 \hspace{0.25mm} C_{qb}^{(8)} + 0.034 \hspace{0.25mm} C_{qtqb}^{(1)} + 0.009 \hspace{0.25mm} C_{qtqb}^{(8)} \,, \hspace{2mm} \\[2mm]
\frac{\delta \kappa_\gamma}{v^2} = 0.023 \hspace{0.25mm} C_{qt}^{(1)} + 0.030 \hspace{0.25mm} C_{qt}^{(8)} - 0.0001 \hspace{0.25mm} C_{qb}^{(1)} - 0.0002 \hspace{0.25mm} C_{qb}^{(8)} - 0.003 \hspace{0.25mm} C_{qtqb}^{(1)} - 0.0003 \hspace{0.25mm} C_{qtqb}^{(8)} \,. \hspace{2mm}
\end{gathered}
\eea
A few comments seem appropriate. First, the process $gg \to h$ is in general more sensitive to the Wilson coefficients of the operators in~(\ref{eq:4HFoperators}) compared to $h \to \gamma \gamma$. This~arises from the fact that, in the SM, the decay rate of $h \to \gamma \gamma$ is dominated by the contribution from the~$W$~boson, which remains unaffected by the operators~(\ref{eq:4HFoperators}) up to the three-loop level. Second, the numerical coefficients multiplying $C_{qb}^{(1)}$ in~(\ref{eq:kappaggammanumerics}) are significantly smaller than those for~$C_{qt}^{(1)}$. This is due to a relative Yukawa suppression of the form $m_b^2/m_t^2 \hspace{0.125mm} \ln^2 \left (m_h^2/m_b^2 \right ) \simeq 1/30$, featuring a Sudakov-like logarithm characteristic of the bottom-quark contributions to the processes $gg \to h$ and $h \to \gamma \gamma$. The Wilson coefficient~$C_{qb}^{(8)}$ exhibits formally the same relative suppression compared to $C_{qt}^{(8)}$. Finally, the first formula in~(\ref{eq:kappaggammanumerics}) shows a pronounced sensitivity to~$C_{qtqb}^{(1)}$. This feature arises from the two-loop form factors $G_{bt}^{(2)}$, where the Wilson coefficient is enhanced by a factor of~$m_t/m_b\simeq 36$. As pointed out in~\cite{Gauld:2015lmb}, in the context of the~$h \to b \bar b$ decay, such an enhancement arises only when no particular UV completion is specified. However, in a broad class of UV completions, such as BSM models with minimal flavor violation~\cite{DAmbrosio:2002vsn}, the Wilson coefficient $C_{qtqb}^{(1)}$ is expected to scale as $y_t \hspace{0.125mm} y_b$, effectively eliminating the~$m_t/m_b$ enhancement in these BSM scenarios. The same arguments apply to the Wilson coefficient $C_{qtqb}^{(8)}$. 

As an illustrative example, we now derive constraints on the Wilson coefficients of the operators in~(\ref{eq:4HFoperators}) using the existing LHC signal strength measurements of Higgs production in gluon-gluon fusion. At the 
$68\%$~confidence level~(CL), the ATLAS collaboration reports the following constraint
\beq \label{eq:kappagATLASRunII}
\kappa_g = 0.949_{-0.067}^{+0.072} \,, 
\eeq
derived from the full LHC~Run~2~dataset~\cite{ATLAS:2022vkf}. Applying~(\ref{eq:kappaggammanumerics}), we obtain the following single parameter bounds on the relevant Wilson coefficients: 
\beq \label{eq:ourlimits}
\begin{gathered}
C_{qt}^{(1)} = \frac{[-4.2, 23]}{{\rm TeV}^2} \,, \qquad C_{qt}^{(8)} = \frac{[-17, 90]}{{\rm TeV}^2} \,, \\[2mm]
C_{qb}^{(1)} = \frac{[-1026, 191]}{{\rm TeV}^2} \,, \qquad C_{qb}^{(8)} = \frac{[-878, 163]}{{\rm TeV}^2} \,, \\[2mm]
C_{qtqb}^{(1)} = \frac{[-57, 11]}{{\rm TeV}^2} \,, \qquad C_{qtqb}^{(8)} = \frac{[-227, 42]}{{\rm TeV}^2} \,. 
\end{gathered}
\eeq
Note that since the expressions in~(\ref{eq:kappaggammanumerics}) use $\mu = m_h$, the above constraints apply to the low-scale Wilson coefficients. Limits on the Wilson coefficients~$C_{qt}^{(1)}$ and~$C_{qt}^{(8)}$ have also been obtained from the global analysis of EW precision observables~\cite{Dawson:2022bxd,DiNoi:2025uhu} and LHC top-quark data~\cite{Ethier:2021bye,Degrande:2024mbg,DiNoi:2025uhu}. The constraints reported in~(\ref{eq:ourlimits}) are found to be weaker than those derived in these studies. At the one-loop level, the~$Z \to b \bar b$ decay also becomes sensitive to $C_{qb}^{(1)}$~\cite{Dawson:2022bxd,Haisch:2024wnw}. Using the results from~\cite{Haisch:2024wnw}, we find that the constraints on this Wilson coefficient from the~$Z$-pole measurements by SLC and LEP~\cite{ALEPH:2005ab} are stronger than those given in~(\ref{eq:ourlimits}). To the best of our knowledge, no constraints on $C_{qb}^{(8)}$ have been derived, making the weak limit obtained here nominally the most stringent. The Wilson coefficients~$C_{qtqb}^{(1)} $ and $C_{qtqb}^{(8)}$ can be probed in the~$h \to b \bar b$ decay~\cite{Gauld:2015lmb}. Based on the most recent signal strength measurements for $h \to b \bar b$ by ATLAS presented in~\cite{ATLAS:2022vkf}, we find that the constraints on $C_{qtqb}^{(1)} $ and $C_{qtqb}^{(8)}$ from Higgs decays to bottom quarks are stronger than those shown in~(\ref{eq:ourlimits}). This~occurs because the operators $Q_{qtqb}^{(1)} $ and $Q_{qtqb}^{(8)}$ affect the~$h \to b \bar b$ decay at the one-loop level, while $gg \to h$ is modified first at the two-loop level. 

Finally, we note that combining the information from the~$gg \to h$ and $h \to \gamma \gamma$ processes enables one to lift the flat directions in the parameter space of $C_{qt}^{(1)}$ and $C_{qt}^{(8)}$ that appear in the expressions for~$\delta \kappa_g$ and $\delta \kappa_\gamma$ as given in~(\ref{eq:kappaggammanumerics}). As very recently emphasized in~\cite{DiNoi:2025uhu}, this feature emerges only when the two-loop matching corrections~(\ref{eq:Gij2}) and~(\ref{eq:Aij2}) are computed using the NDR scheme for~$\gamma_5$. In~contrast, as discussed in Appendix~\ref{app:HVresults}, when working in the~HV~scheme, both~$\delta \kappa_g$ and $\delta \kappa_\gamma$ are, to linear order in the Wilson coefficients, proportional to the combination $C_{qt}^{(1)} + C_F \hspace{0.25mm} C_{qt}^{(8)}$. Consequently, if the computation is performed in this scheme for $\gamma_5$, combining the~$gg \to h$ and $h \to \gamma \gamma$ channels does not resolve the ambiguity.

\section{Conclusions} 
\label{sec:conclusions}

This article can be viewed as a sequel to our recent work~\cite{Haisch:2024wnw}, in which we computed the one- and two-loop SMEFT matching corrections induced by third-generation four-quark operators for EW precision measurements and flavor observables. The insights gained from that paper are used here to calculate the two-loop contributions from the third-generation four-quark operators~(\ref{eq:4HFoperators}) to the~$gg \to h$ and $h \to \gamma \gamma$ processes. Following the discussion in~\cite{Alasfar:2022zyr,DiNoi:2023ygk,Heinrich:2023rsd}, the relevant two-loop amplitudes have been categorized into three distinct topology classes. For each topology, we have provided a brief discussion of the renormalization procedure that ensures the corresponding amplitudes are UV finite. It turns out that, in the NDR scheme, the renormalization of the two-loop amplitudes requires certain two-loop SMEFT anomalous dimensions, which we compute as a byproduct. For a more detailed discussion of the technical aspects of the two-loop calculations of third-generation four-quark operator contributions to Higgs physics, we refer the reader to Appendices~\ref{app:mixren},~\ref{app:MSshifts}, and~\ref{app:HVresults}, as~well~as to the articles~\cite{DiNoi:2023ygk,Heinrich:2023rsd}.

The two-loop beta functions and matching conditions derived in this work are scheme-dependent. The results given in~(\ref{eq:beta2}),~(\ref{eq:Gij2}), and~(\ref{eq:Aij2}) correspond to an NDR treatment of~$\gamma_5$ and a mixed~$\overline{\rm MS\hspace{0.25mm}}$-$\hspace{0.125mm}{\rm OS}$ scheme, with respect to the renormalization of operators and heavy-quark masses and Yukawa couplings, respectively. The observed scheme dependencies are an intrinsic and unavoidable feature of NLO SMEFT calculations that treat the high-scale Wilson coefficients as arbitrary.\footnote{A concise discussion of the regularization and renormalization scheme dependencies of our two-loop results is provided in Appendices~\ref{app:MSshifts} and~\ref{app:HVresults}.} This occurs because the scheme dependence of the high-scale Wilson coefficients, as determined by an explicit matching calculation to a BSM model in a specific scheme, is exactly canceled by the scheme dependence of the anomalous dimensions, resulting in a scheme-independent outcome for all physical observables order by order in perturbation theory. Given these ambiguities, it is clear that NLO SMEFT computations can be simplified by choosing an appropriate regularization and renormalization scheme, along with a well-suited operator basis. Regarding the third-generation four-quark operators~(\ref{eq:4HFoperators}), we believe that employing the NDR scheme in combination with the operator basis introduced by the LHC Top Working Group~(LHCTopWG) in~\cite{Aguilar-Saavedra:2018ksv} is a particularly useful choice. The NDR scheme is advantageous over alternatives like the HV scheme because of its computational efficiency and simplicity. Additionally, as demonstrated in~\cite{Haisch:2024wnw}, the LHCTopWG operator basis offers a key advantage over the Warsaw operator basis, as it avoids non-zero traces involving~$\gamma_5$, which would otherwise require the introduction of Wess-Zumino terms~\cite{Wess:1971yu}. 

We have also touched upon the phenomenological implications of our two-loop calculations. To this purpose, we have derived numerical expressions for the signal strength modifications in $gg \to h$ and $h \to \gamma \gamma$. We found that the process $gg \to h$ is in general more sensitive to the Wilson coefficients of third-generation four-quark operators than $h \to \gamma \gamma$. This is a simple consequence of the fact that, in the SM, the decay rate of $h \to \gamma \gamma$ is dominated by the contribution from $W$-boson loops, which effectively suppresses the impact of~(\ref{eq:4HFoperators}) in the~$h \to \gamma \gamma$ signal strength. The derived formulas were finally used to obtain constraints on the low-scale Wilson coefficients of third-generation four-quark operators from gluon-gluon fusion Higgs production. Overall, we found that the resulting bounds are weaker than those derived from EW precision observables, LHC top-quark data, and Higgs decays to bottom quarks~\cite{Ethier:2021bye,Degrande:2024mbg,Haisch:2024wnw,Dawson:2022bxd,Gauld:2015lmb,DiNoi:2025uhu}. The only exception is $C_{qb}^{(8)}$, for which, to the best of our knowledge, no existing constraints have been established. As a result, the limit derived here, while weak, is nominally the most stringent. Nevertheless, the concise analytical and numerical expressions provided in this study should facilitate the integration of constraints on third-generation four-quark operators from $gg \to h$ and $h \to \gamma \gamma$ into global SMEFT~analyses.

\acknowledgments{UH would like to thank Ramona Gr{\"o}ber for valuable discussions more than two years ago that sparked this work. He is also grateful to Luc~Schnell and Ben~Stefanek for their collaborations on related topics. We also thank an anonymous referee for pointing out that the results for $\beta_{HB}^{(2)}$, $\beta_{HW}^{(2)}$, and $\beta_{HWB}^{(2)}$ given in~(\ref{eq:beta2}) of the original manuscript were not consistent with those of~\cite{Born:2026xkr}, which appeared after our work. We have corrected these results in the revised version.} The~Feynman diagrams shown in this article have been drawn with~{\tt FeynArts}.

\begin{appendix}

\section{OS renormalization}
\label{app:mixren}

The two-loop form factors~(\ref{eq:Gij2}) and~(\ref{eq:Aij2}) are computed within a mixed~$\overline{\mathrm{MS}}$--$\mathrm{OS}$ renormalization framework, where the operators are renormalized in the~$\overline{\mathrm{MS}}$ scheme, while the heavy-quark masses and Yukawa couplings are treated in the~OS~scheme. For completeness, we present below the explicit OS~counterterms for the heavy-quark masses and Yukawa couplings, adopting the NDR scheme for $\gamma_5$.

A direct one-loop calculation of the on-shell amplitudes for $t \to t$ and $b \to b$ yields the following expression for the on-shell counterterm of the heavy-quark masses:
\beq \label{eq:massct}
\begin{split}
\delta m_i & = - \frac{m_i^3}{16 \pi^2} \left (\frac{4}{\epsilon} + 2 + 4 L_i \right ) \left ( C_{qi}^{(1)} + C_F \hspace{0.25mm} C_{qi}^{(8)} \right ) \\[2mm]
& \phantom{xx} + \frac{m_j^3}{16 \pi^2} \left (\frac{1}{\epsilon} + 1 + L_j \right ) \left [ \hspace{0.25mm} \big ( 2 C_A + 1 \big) \hspace{0.25mm} C_{qtqb}^{(1)} + C_F \hspace{0.25mm} C_{qtqb}^{(8)} \right ] \,.
\end{split}
\eeq

The counterterm for the $t \bar t \to h$ and $b \bar b \to h$ decay amplitudes receives several contributions, as discussed, for instance, in~\cite{Gauld:2015lmb}. One contribution arises from the renormalization of the SM tree-level $i \bar i \to h$ vertex proportional to the Yukawa coupling $y_i$. In~the~OS~scheme, the corresponding counterterm is
\beq \label{eq:yukawact}
\delta y_i = y_i \hspace{0.5mm} \frac{\delta m_i}{m_i} \,, 
\eeq
with $\delta m_i$ defined in~(\ref{eq:massct}). We note that the expression in~(\ref{eq:massct}) agrees with~\cite{Gauld:2015lmb}, but differs from the result reported in~\cite{Alasfar:2022zyr}.

\section{$\bm{\overline{\rm MS}}$ shifts}
\label{app:MSshifts}

In the main body of our work the heavy-quark masses and Yukawa couplings were renormalized in the~OS~scheme, employing the NDR prescription for~$\gamma_5$. Below we provide the finite shifts for the two-loop form factors~(\ref{eq:Gij2}) and~(\ref{eq:Aij2}) corresponding to a $\overline{\rm MS}$ renormalization of these quantities, still within the NDR scheme. These shifts must be added to the two-loop form factors to convert them from the~OS~scheme to the~$\overline{\rm MS}$ scheme.

For the two-loop form factors in $gg \to h$, the~$\overline{\rm MS}$ shifts associated with the renormalization of the heavy-quark masses are given by
\bea \label{eq:massGij}
\begin{split}
& \delta_m G_{ii}^{(2)} = \frac{3}{2} \hspace{0.5mm} T_F \hspace{0.25mm} \tau_i \hspace{0.25mm} \Big [ 2 \tau_i - \tau_i \hspace{0.25mm} F_{\rm bub} (x_i) + 2 \left ( 3 \tau_i - 1 \right )F_{\rm tri} (x_i) \Big ] \left(1 + 2 L_i \right ) \left ( C_{qi}^{(1)} + C_F \hspace{0.25mm} C_{qi}^{(8)} \right ) \,, \\[4mm]
& \hspace{1cm} \delta_m G_{ij}^{(2)} = -\frac{3 \hspace{0.25mm} m_j}{4 \hspace{0.25mm} m_i} \hspace{0.5mm} T_F \hspace{0.25mm} \tau_j \hspace{0.25mm} \Big [ 2 \tau_i - \tau_i \hspace{0.25mm} F_{\rm bub} (x_i) + 2 \left ( 3 \tau_i - 1 \right ) F_{\rm tri} (x_i) \Big ]\left ( 1 + 2 L_j \right )\\[2mm] 
& \hspace{2.6cm} \times \left [ \hspace{0.25mm} \big ( 2 C_A + 1 \big) \hspace{0.25mm} C_{qtqb}^{(1)} + C_F \hspace{0.25mm} C_{qtqb}^{(8)} \right ] \,.
\end{split}
\eea
The corresponding $\overline{\rm MS}$ shifts related to the renormalization of the heavy-quark Yukawa couplings take the form
\beq \label{eq:yukawaGij}
\begin{gathered}
\delta_y G_{ii}^{(2)} = \frac{1}{2} \hspace{0.5mm} {\tau_i} \hspace{0.5mm} G_i^{(1)} \left ( 1 + 2 L_i \right ) \left ( C_{qi}^{(1)} + C_F \hspace{0.25mm} C_{qi}^{(8)} \right ) \,, \\[2mm]
\delta_y G_{ij}^{(2)} = -\frac{m_j}{4 \hspace{0.25mm} m_i} \hspace{0.5mm} {\tau_j} \hspace{0.25mm} G_i^{(1)} \left ( 1 + 2 L_j \right ) \left [ \hspace{0.25mm} \big ( 2 C_A + 1 \big) \hspace{0.25mm} C_{qtqb}^{(1)} + C_F \hspace{0.25mm} C_{qtqb}^{(8)} \right ] \,,
\end{gathered}
\eeq
where $G_i^{(1)}$ is the form factor~(\ref{eq:gghoneloop}) induced at the one-loop level in the SM by the virtual exchange of heavy quarks.

In the case of the~$h \to \gamma \gamma$ decay, the corresponding $\overline{\rm MS}$ shifts resulting from the renormalization of the heavy-quark masses are given by
\bea \label{eq:massAij}
\begin{split}
& \delta_m A_{ii}^{(2)} = C_A \hspace{0.125mm} Q_i^2 \hspace{0.5mm} \tau_i \hspace{0.25mm} \Big [ 2 \tau_i - \tau_i \hspace{0.25mm} F_{\rm bub} (x_i) + 2 \left ( 3 \tau_i - 1 \right ) F_{\rm tri} (x_i) \Big ] \left(1 + 2 L_i \right ) \left ( C_{qi}^{(1)} + C_F \hspace{0.25mm} C_{qi}^{(8)} \right ) \,, \\[4mm]
& \hspace{0.75cm} \delta_m A_{ij}^{(2)} = -\frac{m_j}{2 \hspace{0.25mm} m_i} \hspace{0.5mm} C_A \hspace{0.25mm} Q_i^2 \hspace{0.25mm} \tau_j \hspace{0.5mm} \Big [ 2 \hspace{0.125mm} \tau_i - \tau_i \hspace{0.25mm} F_{\rm bub} (x_i) + 2 \left ( 3 \tau_i - 1 \right ) \hspace{0.25mm} F_{\rm tri} (x_i) \Big ]\left ( 1 + 2 L_j \right )\\[2mm] 
& \hspace{2.3cm} \times \left [ \hspace{0.25mm} \big ( 2 C_A + 1 \big) \hspace{0.25mm} C_{qtqb}^{(1)} + C_F \hspace{0.25mm} C_{qtqb}^{(8)} \right ] \,, 
\end{split}
\eea
while the finite contributions arising from the renormalization of the heavy-quark Yukawa couplings in the~$\overline{\rm MS}$ scheme can be expressed as
\beq \label{eq:yukawaAij}
\begin{gathered}
\delta_y A_{ii}^{(2)} = \frac{1}{2} \hspace{0.5mm} {\tau_i} \hspace{0.5mm} A_i^{(1)} \left ( 1 + 2 L_i \right ) \left ( C_{qi}^{(1)} + C_F \hspace{0.25mm} C_{qi}^{(8)} \right ) \,, \\[2mm]
\delta_y A_{ij}^{(2)} = -\frac{m_j}{4 \hspace{0.25mm} m_i} \hspace{0.5mm} {\tau_j} \hspace{0.25mm} A_i^{(1)} \left ( 1 + 2 L_j \right ) \left [ \hspace{0.25mm} \big ( 2 C_A + 1 \big) \hspace{0.25mm} C_{qtqb}^{(1)} + C_F \hspace{0.25mm} C_{qtqb}^{(8)} \right ] \,.
\end{gathered}
\eeq
Here, $A_i^{(1)}$ denotes the form factor~(\ref{eq:1loopffdecay}) that accounts for the one-loop corrections from heavy quarks in the SM.

\section{HV results}
\label{app:HVresults}

The results for the two-loop beta functions~(\ref{eq:beta2}) and form factors~(\ref{eq:Gij2}) and~(\ref{eq:Aij2}) correspond to the NDR scheme, employing a mixed $\overline{\rm MS\hspace{0.25mm}}$–${\rm OS}$ renormalization for the operators and heavy-quark masses and Yukawa couplings, respectively. In this appendix, we discuss how these results are changed if the computations are performed in the HV scheme. 

To investigate the scheme dependencies of our two-loop results related to the choice of $\gamma_5$, we first analyze the~$gg \to h$ process. A simple one-loop computation of the on-shell amplitudes for~$t \bar t \to g$ and $b \bar b \to g$ yields the following result:
\bea \label{eq:penguin}
\begin{gathered}
\includegraphics[width=0.225\textwidth]{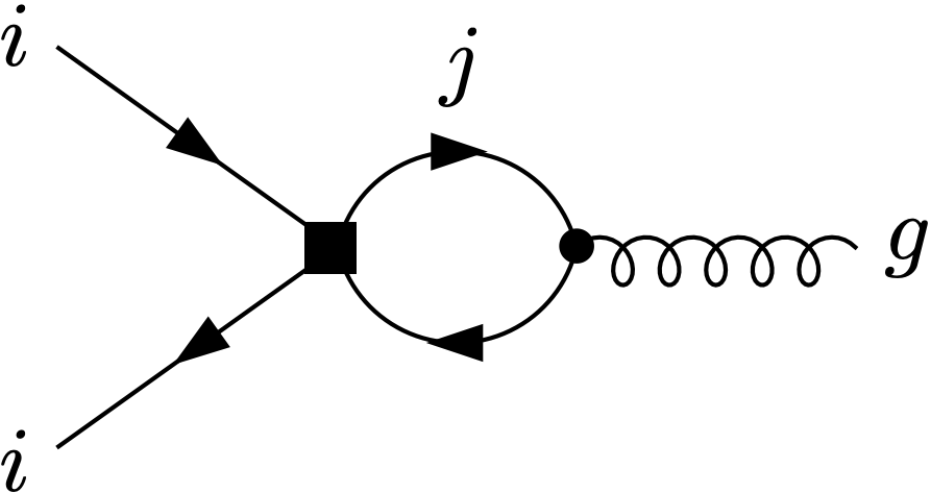} \\[-2mm]
\end{gathered}
\hspace{-3mm} & \displaystyle = \, -\frac{2 g_s T_F }{16 \pi^2} \, \Bigg \{ 2 \hspace{0.25mm} y_i \hspace{0.125mm} K_{\rm scheme} \left [ C_{qi}^{(1)} + \left ( C_F - \frac{C_A}{2} \right ) C_{qi}^{(8)} \right ] \hspace{6mm} \nonumber \\[-4mm] \\[-4mm]
& \displaystyle \hspace{2.65cm} + \frac{y_j}{\epsilon} \left [ C_{qtqb}^{(1)} + \left ( C_F - \frac{C_A}{2} \right ) C_{qtqb}^{(8)} \right ] \Bigg \} \; Q_{iG} \,. \nonumber \hspace{6mm} 
\eea
This demonstrates that certain third-generation four-quark operators~(\ref{eq:4HFoperators}) lead to a non-zero $i \bar i \to g$ matrix element proportional to the chromomagnetic dipole operators $Q_{iG}$ defined in~(\ref{eq:renoperators}). The key point is now that the finite term in~(\ref{eq:penguin}) proportional to $K_{\rm scheme}$ depends on the scheme used for $\gamma_5$, whereas the contribution proportional to $1/\epsilon$ is scheme-independent. Specifically, in agreement with the works~\cite{DiNoi:2023ygk,Heinrich:2023rsd,DiNoi:2024ajj,DiNoi:2025uan,DiNoi:2025uhu}, we find that $K_{\rm NDR} =1$ and $K_{\rm HV} = 0$. In the two-loop calculation of the~$gg \to h$ process, the amplitudes~(\ref{eq:penguin}) appear as one-loop subdiagrams in the third class of graphs depicted in Figure~\ref{fig:diagrams}, where they multiply a UV-divergent triangle loop. This makes these contributions scheme-dependent, as they are proportional to $K_{\rm scheme}$. The same reasoning applies to the on-shell~$t \bar t \to \gamma$ and $b \bar b \to \gamma$ amplitudes, with the small difference that in this case, one must consider the EW dipole operators $Q_{iB}$ and $Q_{iW}$ that appear in~(\ref{eq:renoperators}). From the above, it follows that in the HV scheme, where $K_{\rm HV} = 0$, the two-loop beta functions presented in~(\ref{eq:beta2}) vanish identically. Similarly, the second term in $G_{ii}^{(2)}$ and $A_{ii}^{(2)}$ as given in~(\ref{eq:Gij2}) and~(\ref{eq:Aij2}), respectively, is zero in the HV~scheme. 

A second source of scheme dependence related to $\gamma_5$ originates from the~$t \bar t \to h$, $b \bar b \to h$ and $t \to t$, $b \to b$ amplitudes, which appear as one-loop subdiagrams in the second and first classes of the Feynman diagrams shown in Figure~\ref{fig:diagrams}, respectively.  Computing the on-shell one-loop amplitudes for the processes $t \bar t \to h$ and $b \bar b \to h$, and including the OS counterterm for the corresponding Yukawa coupling given in~(\ref{eq:yukawact}), yields the following result:
\bea \label{eq:giraffe}
\begin{gathered}
\includegraphics[width=0.5\textwidth]{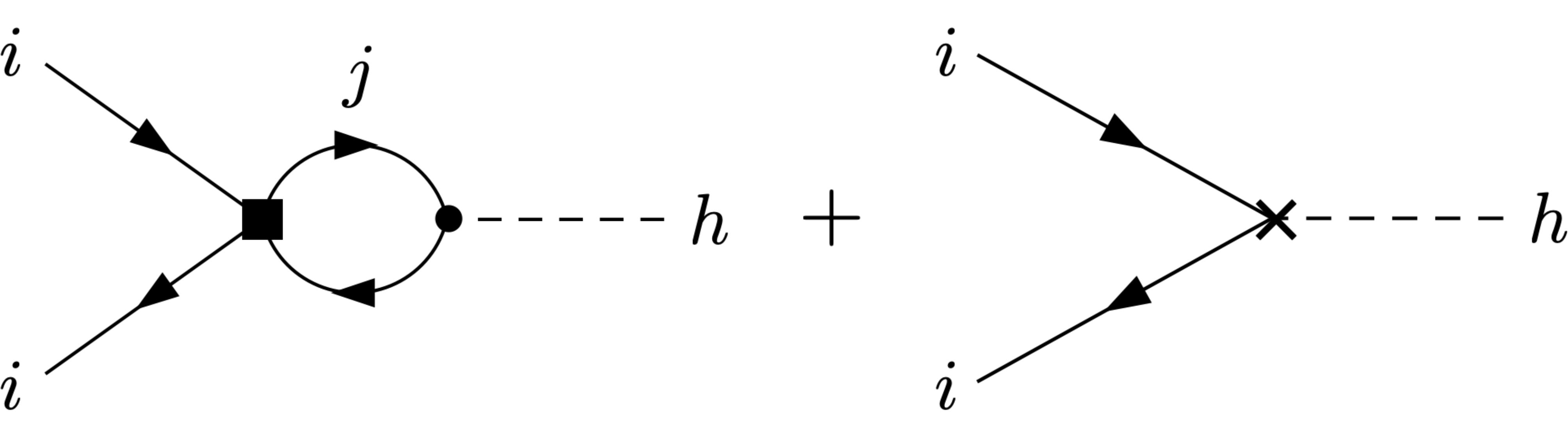} \\[-2mm]
\end{gathered}
\hspace{-6.125cm} & \displaystyle = \nonumber \\[2mm]
\hspace{-8.5cm} & \displaystyle \frac{1}{16 \pi^2} \frac{m_h^2}{v^2} \; \Bigg \{ \left [ \frac{2}{3} \hspace{0.5mm} y_i \left ( \tau_i - 1 \right ) \hspace{0.125mm} K_{\rm scheme} - \frac{4}{3} \hspace{0.25mm} y_i \hspace{0.25mm} J_{i}^{(1)}
\right ] \left [ C_{qi}^{(1)} + \left ( C_F - \frac{C_A}{2} \right ) C_{qi}^{(8)} \right ] \hspace{8mm} \\[2mm]
\hspace{-6.5cm} & \displaystyle + \, \frac{1}{3} \hspace{0.5mm} y_j \hspace{0.25mm} J_{j}^{(1)} \left [ \hspace{0.25mm} \big ( 2 C_A + 1 \big) \hspace{0.25mm} C_{qtqb}^{(1)} + C_F \hspace{0.25mm} C_{qtqb}^{(8)} \right ] \Bigg \} \; Q_{iH} \,. \hspace{12mm} \nonumber 
\eea
Here, $J_{i}^{(1)}$ is given by the following expression:
\beq \label{eq:J1}
J_i^{(1)} = \frac{1}{\epsilon} \hspace{0.25mm} \left ( \tau_i - 1 \right ) + \left ( \tau_i - 1 \right ) \Big ( 1 + F_{\rm bub} (x_i) + L_i \Big ) \,, 
\eeq
As before, we have indicated the scheme-dependent terms by $K_{\rm scheme}$, with $K_{\rm NDR} = 1$ and $K_{\rm HV} = 0$ accordingly. The results~(\ref{eq:giraffe}) and (\ref{eq:J1}) are consistent with the findings of the articles~\cite{DiNoi:2023ygk,Heinrich:2023rsd,DiNoi:2024ajj,DiNoi:2025uan,DiNoi:2025uhu}, once one takes into account that some of these studies use the~$\overline{\rm MS}$~scheme, rather than the~OS~scheme, for the renormalization of the Yukawa couplings. In~the two-loop calculation of $gg \to h$ and $h \to \gamma \gamma$, the amplitudes given in~(\ref{eq:giraffe}) appear as one-loop subdiagrams in the second type of graphs illustrated in Figure~\ref{fig:diagrams}, where they are multiplied by a UV-finite triangle loop. As a result, the scheme-dependent term in~(\ref{eq:giraffe}) can be readily incorporated into the expressions~(\ref{eq:Gij2}) and~(\ref{eq:Aij2}) by replacing the bracketed factor $\left(1 + 2 F_{\rm bub}(x_i) + 2 L_i \right)$ with $\left(2 + 2 F_{\rm bub}(x_i) + 2 L_i \right)$, which appears in the first line of both $G_{ii}^{(2)}$ and $A_{ii}^{(2)}$. Here, the first expression corresponds to the NDR scheme, whereas the second pertains to the HV scheme. 

We furthermore note that the one-loop amplitudes for $t \to t$ and $b \to b$ are likewise scheme-dependent. However, this scheme dependence does not impact the two-loop matching conditions presented in Section~\ref{sec:matching}, as long as the heavy-quark masses are renormalized in the~OS~scheme, as implemented in the main text and detailed in~Appendix~\ref{app:mixren}. Finally, it is worth emphasizing that only the contributions in~(\ref{eq:penguin}) and~(\ref{eq:giraffe}) arising from the vector operators~$Q_{qi}^{(1)}$ and~$Q_{qi}^{(8)}$ are sensitive to the specific prescription used for handling~$\gamma_5$, whereas the scalar operators~$Q_{qtqb}^{(1)}$ and $Q_{qtqb}^{(8)}$ yield scheme-independent results. This point should be kept in mind when interpreting the results presented in~Sections~\ref{sec:ADMs}~and~\ref{sec:matching}.

\end{appendix}



%

\end{document}